\documentclass[twocolumn,showpacs,aps,prl,superscriptaddress]{revtex4-1}
\usepackage{ulem,bm}
\usepackage{array,epsfig,amsmath,amssymb}
\usepackage{graphicx} 
\usepackage{dsfont}
\usepackage{color}

\def\bra#1{\langle #1|}
\def\ket#1{|#1 \rangle}

\begin{document}

\title{
Nearly Deterministic Bell Measurement for 
Multiphoton Qubits and Its Application to Quantum Information Processing}

\author{Seung-Woo Lee}
\affiliation{Center for Macroscopic Quantum Control, Department of
Physics and Astronomy, Seoul National University, Seoul, 151-742,
Korea}

\author{Kimin Park}
\affiliation{Center for Macroscopic Quantum Control, Department of
Physics and Astronomy, Seoul National University, Seoul, 151-742,
Korea}

\affiliation{Department of Optics, Palack\'y University, 17. listopadu 1192/12, 77146 Olomouc, Czech Republic}

\author{Timothy C. Ralph}
\affiliation{Centre for Quantum Computation and Communication Technology, School of Mathematics and Physics, University of Queensland, St Lucia, Queensland 4072, Australia}

\author{Hyunseok Jeong}\email{h.jeong37@gmail.com}
\affiliation{Center for Macroscopic Quantum Control, Department of
Physics and Astronomy, Seoul National University, Seoul, 151-742,
Korea}

\begin{abstract}
We propose a Bell measurement scheme by employing a logical qubit in Greenberger-Horne-Zeilinger (GHZ) entanglement with an arbitrary number of photons. Remarkably, the success probability of the Bell measurement as well as teleportation of the GHZ entanglement can be made arbitrarily high using only  linear optics elements and photon on-off measurements as the number of photons increases. Our scheme outperforms previous proposals using single photon qubits  when comparing the success probabilities in terms of the average photon usages. It has another important advantage  for experimental feasibility that it does not require photon number resolving measurements.
Our proposal provides an alternative candidate for all-optical quantum information processing.
\end{abstract}

\maketitle

Photons are a promising candidate for quantum information processing \cite{PKok2007,Ralph2010}.
A well-known method to construct a photonic qubit is to use a single photon with its polarization degree of freedom \cite{PKok2007}. A crucial element in quantum communication and computation using linear optics and photon measurements \cite{Knill2001} is the Bell state measurement that discriminates between four Bell states. The standard Bell-measurement scheme 
for the Bell states of single-photon qubits utilizes beam splitters and photodetectors \cite{Lut99,Calsa2001}.
This method, in effect, projects two photons onto a complete measurement basis of two Bell states and two product states so that only two of the Bell states can be unambiguously identified.
Due to this reason, the success probability of the Bell measurement using linear optics elements and photodetectors is limited to 1/2 \cite{Lut99,Calsa2001}.
This has been a fundamental hindrance to deterministic quantum teleportation and  scalable quantum computation \cite{PKok2007,Ralph2010}.
 There are proposals to improve the success probability of the Bell discrimination using ancillary states \cite{Grice2011,Ewert2014}, additional squeezing operations \cite{Zaidi2013} and different types of qubits encoding using coherent states \cite{Jeong2001} or hybrid states \cite{SWLEE13}.
In fact, all these schemes suffer from the requirement of photon number resolving detection 
\cite{Grice2011,Zaidi2013,Ewert2014,Jeong2001,SWLEE13}. The requirement of ancillary resource entanglement \cite{Grice2011,Ewert2014}  and the limited success probabilities \cite{Zaidi2013} are other features to overcome.

In this paper, we propose a Bell measurement scheme using linear optics and photon on-off measurements with qubit encoding in the form of Greenberger-Horne-Zeilinger (GHZ) entanglement. 
It is shown that the logical Bell states can be efficiently discriminated by performing $N$ times of Bell measurements on the individual photon pairs, where $N$ is the number of photons in a logical qubit,
using only the standard technique with beam splitters and on-off photodetectors.
The limitation that each measurement for photon
pairs can only identify two of the four Bell states is overcome by the
fact that each of the four $N$-photon Bell states is characterized by
the number of contributions from the two single-photon-qubit Bell states that can be
identified in the measurement of photon pairs.
As a result, the logical Bell measurement fails only when none of the $N$ pairs is a detectable Bell state, resulting in a success probability of $1-2^{-N}$ that rapidly approaches unity as $N$ increases;
it outperforms the previous approaches \cite{Grice2011,Zaidi2013,Ewert2014} in its efficiency against the number of photons {\it without} using photon number resolving detection.
Using this Bell measurement scheme, a qubit in an $N$ photon GHZ-type entanglement can be teleported with an arbitrarily high success probability
with a GHZ-type entangled channel of a $2N$ number of photons as $N$ becomes large.
In our framework, a universal set of gate operations can be constructed using only linear optics, on-off measurements and multiphoton entanglement. This may be a competitive new approach to photonic quantum information processing due to the aforementioned advantages.

{\em Multiphoton Bell measurement.--}
We define single-photon-qubit Bell states as
\begin{equation}
\label{eq:BellB}
\begin{aligned}
\ket{\Phi^\pm}=\frac{1}{\sqrt{2}}(\ket{+}\ket{+}\pm\ket{-}\ket{-}),\\
\ket{\Psi^\pm}=\frac{1}{\sqrt{2}}(\ket{+}\ket{-}\pm\ket{-}\ket{+}),
\end{aligned}
\end{equation}
in the diagonal basis $\ket{\pm}=(\ket{H}\pm\ket{V})/\sqrt{2}$ in terms of horizontal and vertical polarization single photon states $\ket{H}$ and $\ket{V}$.
Only two of the four Bell states in Eq.~(\ref{eq:BellB}) can be discriminated 
by the standard Bell measurement technique using linear optics \cite{Lut99,Calsa2001}.
For example, one can identify $\ket{\Phi^{-}}$ and $\ket{\Psi^{-}}$ using beam splitters and four on-off photodetectors \cite{Calsa2001}. We shall refer to this single-photon-qubit Bell measurement  as $\rm B_s$.

\begin{figure}[tp]
\centering
\includegraphics[width=3.4in]{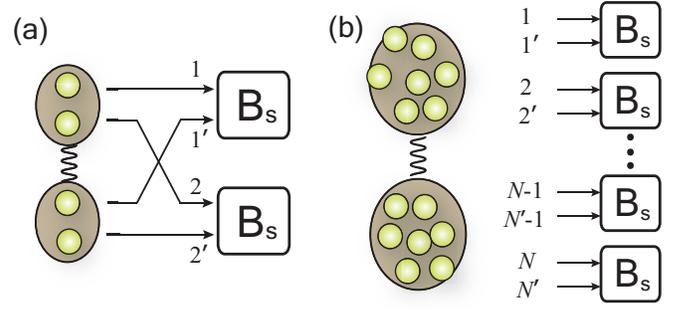}
\caption{(Color online). 
(a) Bell measurement for two-photon qubits using two single-photon-qubit Bell measurements $\rm B_s$. Each logical qubit is of two photons. (b) Bell measurement for $N$-photon qubits through $N$ times of $\rm B_s$ measurements. }\label{fig:scheme}
\end{figure}

The logical basis is defined with $N$ photons as
\begin{equation}
\begin{aligned}
& \ket{0_{L}}\equiv\ket{+}^{\otimes N}=\ket{+}_1\ket{+}_2\ket{+}_3 \cdot\cdot\cdot \ket{+}_N,\\
& \ket{1_{L}}\equiv\ket{-}^{\otimes N}=\ket{-}_1\ket{-}_2\ket{-}_3 \cdot\cdot\cdot \ket{-}_N
 \end{aligned}
\end{equation}
and then a logical qubit is generally in a GHZ-type state as $\alpha |+\rangle^{\otimes N} + \beta |-\rangle^{\otimes N}$.
 Let us first consider the simplest case of two-photon encoding ($N=2$) with $\ket{0}_L\equiv\ket{+}\otimes\ket{+}$ and $\ket{1}_L\equiv\ket{-}\otimes\ket{-}$. The logical Bell states 
can be expressed as
\begin{equation}
\begin{aligned}
&\ket{\Phi^{\pm}_{(2)}}=\frac{1}{\sqrt{2}}(\ket{+}_{1}\ket{+}_{2}\ket{+}_{1'}\ket{+}_{2'}
\pm\ket{-}_{1}\ket{-}_{2}\ket{-}_{1'}\ket{-}_{2'}),\\
&\ket{\Psi^{\pm}_{(2)}}=\frac{1}{\sqrt{2}}(\ket{+}_{1}\ket{+}_{2}\ket{-}_{1'}\ket{-}_{2'}
\pm\ket{-}_{1}\ket{-}_{2}\ket{+}_{1'}\ket{+}_{2'}),
\end{aligned}
\end{equation}
where the first logical qubit is of photonic modes $1$ and $2$ while the second is of $1'$ and $2'$.
Simply by rearranging modes $1'$ and $2$ as implied in Fig.~1(a),
these Bell states can be represented in terms of  the single-photon-qubit Bell states in Eq.~(\ref{eq:BellB}) as
\begin{equation}
\begin{aligned}
&\ket{\Phi^{\pm}_{(2)}}=\frac{1}{\sqrt{2}}(\ket{\Phi^+}_{11'}\ket{\Phi^{\pm}}_{22'}+\ket{\Phi^-}_{11'}\ket{\Phi^{\mp}}_{22'}),\\
&\ket{\Psi^{\pm}_{(2)}}=\frac{1}{\sqrt{2}}(\ket{\Psi^+}_{11'}\ket{\Psi^{\pm}}_{22'}+\ket{\Psi^-}_{11'}\ket{\Psi^{\mp}}_{22'}).
\end{aligned}
\end{equation}
It then becomes clear that the four Bell states $\ket{\Phi^{\pm}_{(2)}}$ and $\ket{\Psi^{\pm}_{(2)}}$ can be discriminated with a 75\% success probability by means of two separate $\rm B_s$ measurements performed on two photons, one from the first qubit and the other from the second as shown in Fig.\ref{fig:scheme}(a). 
Note that  a $\rm B_s$ measurement can identify only $\ket{\Phi^{-}}$ and $\ket{\Psi^{-}}$  with the total success probability 50\%.
From the results of two $\rm B_s$ measurements, one can distinguish the Bell states as follows: (i) $\ket{\Phi^{+}_{(2)}}$ when both $\rm B_s$ measurements succeed with results $\ket{\Phi^-}$, (ii) $\ket{\Phi^{-}_{(2)}}$ when one measurement succeeds with $\ket{\Phi^{-}}$, (iii) $\ket{\Psi^{+}_{(2)}}$ when both succeeds with $\ket{\Psi^{-}}$, (iv) $\ket{\Psi^{-}_{(2)}}$ when one measurement succeeds with $\ket{\Psi^{-}}$, (v) failure occurs when both the measurements fail ({\em i.e.} neither $\ket{\Phi^{-}}$ nor $\ket{\Psi^{-}}$ is obtained). Assuming equal input probabilities of Bell states, we can obtain the success probability of the Bell measurement as $P_s=3/4$.

This scheme can be generalized to arbitrary $N$ photon encoding.
 The logical Bell states $\ket{\Phi^{\pm}_{(N)}}=(|0_L\rangle|0_L\rangle\pm|1_L\rangle|1_L\rangle)/\sqrt{2}$ and $\ket{\Psi^{\pm}_{(N)}}=(|0_L\rangle|1_L\rangle\pm|1_L\rangle|0_L\rangle)/\sqrt{2}$  can be expressed as
\begin{equation}
\begin{aligned}
&\ket{\Phi^{+}_{(N)}}=\frac{1}{\sqrt{2^{N-1}}}\sum^{[N/2]}_{j=0}{\cal P}[\ket{\Phi^+}^{\otimes N-2j}\ket{\Phi^-}^{\otimes 2j})],\\
&\ket{\Phi^{-}_{(N)}}=\frac{1}{\sqrt{2^{N-1}}}\sum^{[(N-1)/2]}_{j=0}{\cal P}[\ket{\Phi^+}^{\otimes N-2j-1}\ket{\Phi^-}^{\otimes 2j+1})],\\
&\ket{\Psi^{+}_{(N)}}=\frac{1}{\sqrt{2^{N-1}}}\sum^{[N/2]}_{j=0}{\cal P}[\ket{\Psi^+}^{\otimes N-2j}\ket{\Psi^-}^{\otimes 2j})],\\
&\ket{\Psi^{-}_{(N)}}=\frac{1}{\sqrt{2^{N-1}}}\sum^{[(N-1)/2]}_{j=0}{\cal P}[\ket{\Psi^+}^{\otimes N-2j-1}\ket{\Psi^-}^{\otimes 2j+1})],
\end{aligned}
\end{equation}
where $[x]$ denotes the maximal integer $\leq x$, and ${\cal P}[\cdot]$ performs the permutation of $N$ elements of photon pairs (Supplementary Material). For example, $\ket{\Phi^{+}_{(3)}}=(\ket{\Phi^{+}}^{\otimes3}+{\cal P}[\ket{\Phi^{+}}\ket{\Phi^{-}}^{\otimes2}])/2 = (\ket{\Phi^{+}}^{\otimes3}+\ket{\Phi^{+}}\ket{\Phi^{-}}^{\otimes2}+\ket{\Phi^{-}}\ket{\Phi^{+}}\ket{\Phi^{-}}+\ket{\Phi^{-}}^{\otimes2}\ket{\Phi^{+}})/2$. The four logical Bell states can be discriminated by performing $N$ times of $\rm B_s$ measurements as illustrated in Fig.~\ref{fig:scheme}(b). Each $\rm B_s$ is performed on two photons, one from the first logical qubit and the other from the second. Clearly, the results of the logical Bell measurement are: (i) $\ket{\Phi^{+}_{(N)}}$ when an even number of $\rm B_s$ measurements succeed with result $\ket{\Phi^-}$, (ii) $\ket{\Phi^{-}_{(N)}}$ for an odd number of $\ket{\Phi^-}$, (iii) $\ket{\Psi^{+}_{(N)}}$ for an even number of $\ket{\Psi^-}$, (iv) $\ket{\Psi^{-}_{(N)}}$ for an odd number of $\ket{\Psi^-}$, (v) the measurement fails when none of the $\rm B_s$ measurements succeeds.
In fact, one can perform the logical Bell measurement effectively via either spatially or temporally distributed $N$ times $\rm B_s$ measurements, irrespectively of the order of measurements. 

%

Assuming equal input probabilities of the Bell states, we can obtain the success probability of the Bell measurement as $P_s=1-2^{-N}$. Remarkably, our scheme shows the best performance among the Bell discrimination schemes for photons with respect to the attained success probability against the average photon number ($\bar{n}$) used in the process as shown in Fig.~2 (details are presented in Supplementary Material). For example, it reaches $P_s=0.996$ with $N=8$ $(\bar{n}=16)$. 
Our scheme does not require photon number resolving detectors in contrast to previous schemes suggested to improve the success probability of a Bell measurement  \cite{Grice2011,Zaidi2013,Ewert2014}.

{\em Nearly deterministic quantum teleportation.--} 
Our Bell measurement scheme 
immediately enhances the success probability of the standard quantum teleportation \cite{Bennett93}. Suppose that an unknown qubit $\ket{\phi_N}_A=a\ket{+}^{\otimes N}_A+b\ket{-}^{\otimes N}_A$ with $N$ photons at site $A$ is to be teleported via a channel state $|+\rangle_A^{\otimes N}|+\rangle_B^{\otimes N}+|-\rangle_A^{\otimes N}|-\rangle_B^{\otimes N}$ to site $B$. 
The sender carries out $N$-times of $\rm B_s$ measurements, 
where each $\rm B_s$ is performed on two photons, i.e., one from $\ket{\phi_N}_A$ and the other from site $A$ of the channel.
The receiver at site $B$ can then retrieve $\ket{\phi_N}$ by performing appropriate unitary transforms. 
The required Pauli X (bit flip) and Z (phase flip) operations in the logical qubit basis can be implemented deterministically by phase-flipping all photon modes and by executing a bit-flip on any one mode, in the $\{|H\rangle,~|V\rangle\}$ baisis, respectively.
Therefore, the success probability of teleportation equals that of the Bell measurement $P_s=1-2^{-N}$.

\begin{figure}[t]
\centering
\includegraphics[width=3.0in]{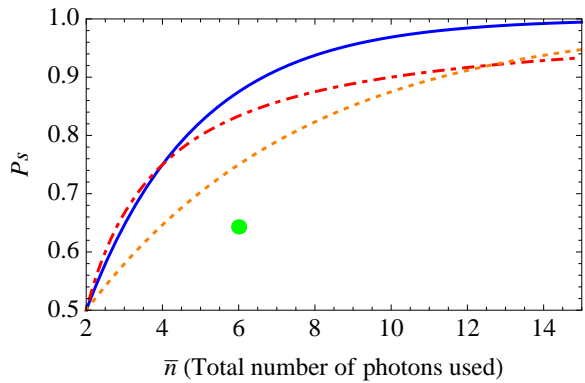}
\caption{(Color online). 
The success probability of Bell measurements against the average photon number ($\bar{n}$) used in the process. It is given as $1-2^{-\bar{n}/2}$ for our Bell measurement scheme (blue curve), $1-1/\bar{n}$ for Grice's scheme (red dot-dashed) \cite{Grice2011}, $P_s=0.643$ with $\bar{n}=6.00029$ for the squeezing scheme (green circle) \cite{Zaidi2013}, and $1-2^{-\bar{n}/4}$ for scheme using ancillary photons (orange dotted) \cite{Ewert2014}.
}\label{fig:tele}
\end{figure}

{\em Universal quantum computation.--} 
Using our framework, 
a universal set of gate operations can be constructed.
For example, Pauli X, arbitrary Z (phase), Hadamard, and a controlled-Z operations constitute such a universal set.
Pauli X and arbitrary Z (phase) operations are straightforward to implement in the way explained earlier for teleportation.
Hadamard and CZ gates can be implemented through the gate teleportation protocol 
with specific types of entangled states \cite{Gottesman1999}.
The success probability of the gate operations based on the teleportation protocol can be made nearly deterministic
by increasing the number of photons for a logical qubit. 
The cost is preparation of mutiphoton entanglement as resource states. 
Such multiphoton entanglement have been experimentally demonstrated \cite{JWPan2012}. For example, GHZ-type entanglement up to 8 photons \cite{Yao2012,Huang2011} and cluster states up to 8 photons \cite{Yao2012Nature} were generated.
On-demand generation schemes \cite{deterministic,Lindener2009} are also expected to be realized based on semiconductor quantum dots \cite{Young2006}.

{\em Effects of photon losses.--} Photon loss is a major detrimental factor in optical quantum information processing \cite{Ralph2010}. 
We assume that the photon loss rate for any single mode is $\eta$  and analyze the errors caused by the photon losses using the master equation (Supplementary Material) \cite{Phoenix90}.
Photon loss during quantum computing occurs with rate $P=1-(1-\eta)^N$ for a logical qubit. If a photon is lost, the qubit experiences a Pauli Z error with probability 1/2. The failure probability ($1-P_{s} $) of the logical Bell measurement is obtained as
\begin{equation}
P_f(\eta)=\sum^{N}_{k=0}\binom{N}{k}(1-\eta)^{N-k}\eta^{k}\left(\frac{1}{2}\right)^{N-k}=\left(\frac{1+\eta}{2}\right)^N,
\end{equation}
where $\binom{N}{k}$ represents the binomial coefficient. Note that errors caused by loss at any single photon mode are in fact detectable by loss of the photon at any detector(s) during the logical Bell measurement. Such an error noticed immediately by a measurement is called ``locatable'' \cite{Ralph2010}. Moreover, missing photons in the input qubit can be compensated at the output qubit as far as the teleportation succeeds. In our scheme, unlocatable errors that should be corrected by an error correction code appear only in quantum memory with rate $P$, i.e., the photon loss rate of a logical qubit. 

\begin{table}[t]
\begin{tabular}{|c|c||c|c|}
\hline
~$N$~~&~Noise threshold $\eta$~~&~$N$~~&~Noise threshold $\eta$~~\\
\hline
~3~~&~$1.3\times 10^{-3}$~~&~6~~&~$1.3\times 10^{-3}$~~\\
\hline
~4~~&~$1.7\times 10^{-3}$~~&~7~~&~$1.1\times 10^{-3}$~~\\
\hline
~5~~&~$1.5\times 10^{-3}$~~&~8~~&~$0.9\times 10^{-3}$~~\\
 \hline
\end{tabular}
\caption{\label{tab:table1}{Fault-tolerant noise thresholds ($\eta$) for different number of photons in a logical qubit ($N$) using the seven-qubit Steane code and the telecorrection protocol \cite{Dawson2006}. The highest threshold is obtained when $N=4$. }}
\end{table}

We summarize the assumptions made for our analysis of quantum computing as follows. Multiphoton entangled states, both for logical qubits and for entangled channels for gate teleportation, are provided by off-line processes. During the off-line process of producing multiphoton entanglement used as quantum channels, loss occurs with rate $\eta$; as a result, imperfect channels (in which photons are lost with rate $\eta$ at each photonic mode) are supplied into the in-line computation process. The initial logical qubits are assumed to be in ideal pure states when they are first supplied into the in-line computation process. During the in-line process of quantum computing, for each gate operation and corresponding time in quantum memory, the same loss rate $\eta$ is applied to each mode of the multiphoton qubits. We note that the total resource cost depends upon the efficiency of the off-line generation process.

{\em Fault-tolerant quantum computation.--} In order to build arbitrary large-scale quantum computers,
the amount of noise per operation with appropriate error corrections should be below a fault-tolerance threshold \cite{Shor1996}. 
We carried out numerical simulations to obtain the threshold for a given loss rate $\eta$. We here employ the seven qubit Steane code \cite{Steane1996} with several levels of concatenation based on the circuit-based telecorrection \cite{Dawson2006}. 
In fact, the Steane code can correct arbitrary logical or unlocatable errors, however for the purpose of this calculation we assume that the errors other than loss errors are negligible compared to the loss errors.
The details of the method \cite{Lund2008,Dawson2006} is presented in the Supplementary Material and the noise thresholds of our model are obtained as shown in Table~\ref{tab:table1}.
Interestingly, the largest threshold is obtained when the qubit is encoded with $4$ photons ($N=4$), and further increase of $N$ lowers the threshold due to the increase of unlocatable errors. The obtained noise threshold ($\sim 1.7\times10^{-3}$) is much higher than those for coherent-state qubits ($\sim2\times10^{-4}$) \cite{Lund2008,coherent_1,coherent_2} and hybrid qubits ($\sim5\times10^{-4}$) \cite{SWLEE13}, and is almost equivalent to the one using parity states \cite{Ralph2005,Hayes2010}. We expect that even much higher thresholds may be attainable by employing recently proposed topological error codes \cite{Yao2012Nature,Sean2010}, which will be interesting future work.

{\em Remarks.--} We have proposed a nearly deterministic Bell discrimination scheme using multiphoton qubit encoding. 
The limitation that only two of four Bell states can be identified by the standard single-photon-qubit Bell measurement, $\rm B_s$, is overcome by multiphoton encoding with GHZ entanglement and $N$ times of $\rm B_s$ measurements, where $N$ is the number of photons in a logical qubit. The logical Bell measurement is performed through $N$ times of $\rm B_s$ measurements and the process  fails only when none of those $N$ times of $\rm B_s$ measurements succeeds. As a result, the  success probability of the logical Bell measurement, $1-2^{-N}$, rapidly approaches to unity as $N$ increases.
It outperforms previous schemes devised to improve success probabilities of Bell measurements using single photons and linear optics, regarding the efficiency in terms of average photon usages. Another remarkable advantage of our scheme over the previous ones is that it does not require photon number resolving measurements but only on-off measurements suffice. It means that all errors due to photon losses are locatable and are relatively easy to handle during quantum information processing. We have  finally demonstrated fault-tolerant quantum computation using our approach. Remarkably, the highest noise-threshold is obtained with $4$-photon qubits and $8$-photon entangled channels that are accessible  in current laboratories \cite{Yao2012,Huang2011,Yao2012Nature}.

Our scheme for the Bell measurement can be performed via either spatially or temporally distributed $N$ times $\rm B_s$ measurements. 
We note that such an experiment can be performed utilizing temporal mode entanglement as done in Refs.~\cite{ex1,ex2,ex3}. It then follows that only one single-photon-qubit Bell-measurement device \cite{Lut99}  is sufficient to perform temporally separate $N$ number of $\rm B_s$ measurements for a logical Bell measurement.
As a proof-of-principle experiment of our scheme, quantum teleportation from two transmitters to two receivers using four-photon entanglement and two $\rm B_s$ measurements, for example, would be immediately realizable  using current technology. 
Our idea, in principle, is not limited to optical systems but can be applied to other multipartite systems. It reveals the possibility of using multipartite entangled systems for efficient quantum communication and computation.

We thank Casey Myers for  useful discussions. This work was supported by the National Research
Foundation of Korea (NRF) grant funded by the Korea government (MSIP) (No. 2010-0018295)
and by the Australian Research Council Centre of Excellence for
Quantum Computation and Communication Technology
(Project No. CE11000102). K.P. acknowledges financing by the European Social Fund and
the state budget of the Czech Republic, POST-UP NO
CZ.1.07/2.3.00/30.0004.


\pagebreak
\begin{center}
\textbf{\large Supplemental Material}
\end{center}
\setcounter{equation}{0}
\setcounter{figure}{0}
\setcounter{table}{0}
\renewcommand{\theequation}{S\arabic{equation}}
\renewcommand{\thefigure}{S\arabic{figure}}
\renewcommand{\bibnumfmt}[1]{[S#1]}
\renewcommand{\citenumfont}[1]{S#1}
\section{Multipartite Bell measurement for arbitrary $N$ photons}

The Bell states with arbitrary $N$ photons given in Eq.~(4) of the main Letter can be represented by single-photon-qubit Bell states. From Eq.~(1) of the main text, we find $\ket{\pm}\ket{\pm}=(\ket{\Phi^{+}}\pm\ket{\Phi^{-}})/\sqrt{2}$ and $\ket{\pm}\ket{\mp}=(\ket{\Psi^{+}}\pm\ket{\Psi^{-}})/\sqrt{2}$. Thus, the Bell states with $N$ photons can be written as
\begin{equation}
\nonumber
\begin{aligned}
&\ket{\Phi^{\pm}_{(N)}}=\frac{1}{\sqrt{2}}(\ket{+}^{\otimes N}\ket{+}^{\otimes N}\pm\ket{-}^{\otimes N}\ket{-}^{\otimes N})\\
&=\frac{1}{\sqrt{2}}\Big((\ket{+}\ket{+})^{\otimes N}\pm(\ket{-}\ket{-})^{\otimes N}\Big)\\
&=\frac{1}{\sqrt{2^{N+1}}}\Big((\ket{\Phi^{+}}+\ket{\Phi^{-}})^{\otimes N}\pm(\ket{\Phi^{+}}-\ket{\Phi^{-}})^{\otimes N}\Big),\\
&\ket{\Psi^{\pm}_{(N)}}=\frac{1}{\sqrt{2}}(\ket{+}^{\otimes N}\ket{-}^{\otimes N}\pm\ket{-}^{\otimes N}\ket{+}^{\otimes N})\\
&=\frac{1}{\sqrt{2}}\Big((\ket{+}\ket{-})^{\otimes N}\pm(\ket{-}\ket{+})^{\otimes N}\Big)\\
&=\frac{1}{\sqrt{2^{N+1}}}\Big((\ket{\Psi^{+}}+\ket{\Psi^{-}})^{\otimes N}\pm(\ket{\Psi^{+}}-\ket{\Psi^{-}})^{\otimes N}\Big),
\end{aligned}
\end{equation}
from which we can obtain Eq.~(5) in the main Letter. For example, we have
\begin{equation}
\nonumber
\begin{aligned}
&\ket{\Phi^{+}_{(3)}}=\frac{1}{4}\Big((\ket{\Phi^{+}}+\ket{\Phi^{-}})^{\otimes 3}+(\ket{\Phi^{+}}-\ket{\Phi^{-}})^{\otimes 3})\Big)\\
&=\frac{1}{2}\Big(\ket{\Phi^{+}}\ket{\Phi^{+}}\ket{\Phi^{+}}+\ket{\Phi^{+}}\ket{\Phi^{-}}\ket{\Phi^{-}}\\
&~~~~~~~~+\ket{\Phi^{-}}\ket{\Phi^{+}}\ket{\Phi^{-}}+\ket{\Phi^{-}}\ket{\Phi^{-}}\ket{\Phi^{+}})\Big)\\
&=\frac{1}{2}\Big(\ket{\Phi^{+}}^{\otimes3}+{\cal P}[\ket{\Phi^{+}}\ket{\Phi^{-}}^{\otimes2}]\Big),
\end{aligned}
\end{equation}
where ${\cal P}[\cdot]$ is the permutation function defined in the main Letter. Likewise, all other Bell states with arbitrary $N$ can be represented in this way.

\section{Comparison with other Bell measurement schemes}

We compare the efficiency of our scheme with those of other recent proposals using single photons \cite{sGrice2011,sZaidi2013,sEwert2014}. In order to make a fair comparison, we consider the success probability of the Bell discrimination in terms of the total number of photons ($\bar{n}$) used in the process. The photons contained in both the logical qubits to be measured as well as the ancillary systems used in the process are counted in $\bar{n}$. We also consider the increase of the average photon number by the squeezing operation when it is used.

(i) In our scheme, total $2N$ photons are used for the Bell measurement because each logical qubit is constructed with $N$ photons and no additional photons are necessary in the process. Thus, the success probability $P_s=1-2^{-\bar{n}/2}$ is achieved with total $\bar{n}=2N$ photons.

(ii) In Grice's proposal~\cite{sGrice2011}, it was shown that the success probability $P_s=1-2^{-N_a}$ is reachable using  $2^{N_a}-2$ ancillary photons. Here $N_a$ is a parameter to denote the ancillary entangled states $|\Gamma_j\rangle$ with $j=1,...,N_a-1$, each of which contains $2^{j}$ photons. The total number of photon usage can thus be obtained by counting all photons in two qubits and ancillary states as $\bar{n}=2+2^{N_a}-2=2^{N_a}$, by which we can rewrite the success probability as $P_s=1-1/\bar{n}$.

(iii) In Zaidi and van Loock's proposal \cite{sZaidi2013}, each mode of Bell states is squeezed by the squeezing operator $S(r)=\exp[-r(a^{\dagger 2}-a^{2})/2]$ with squeezing parameter $r$ to increase the success probability of Bell discrimination. Two indistinguishable Bell states (here assumed as $\ket{\phi^{\pm}}=2^{-1/2}(\ket{HH}\pm\ket{VV})$), passing through a beam splitter as $U_\mathrm{BS}\ket{\phi^{\pm}}$ and squeezed, can be written in the dual-rail representation as
\begin{equation}
\nonumber
\frac{i}{2}(\ket{2'0'0'0'}+\ket{0'0'0'2'}\pm\ket{0'2'0'0'}\pm\ket{0'0'2'0'})
\end{equation}
where $\ket{n'}=S(r)\ket{n}$. 
The average photon number $\bar{n}$ in all four modes can be obtained by
\begin{equation}
\nonumber 
\bra{\phi^{\pm}}S_{1,2,3,4}(-r)(\hat{n}_1+\hat{n}_2+\hat{n}_3+\hat{n}_4)S_{1,2,3,4}(r)\ket{\phi^{\pm}}
\end{equation}
where $\hat{n}_i$ is the photon number operator in the $i$-th mode and $S_{1,2,3,4}(r)=S_1(r)S_2(r)S_3(r)S_4(r)$. Using the relation
\begin{equation}
\nonumber
S(-r)\hat{n}S(r)=(a^\dagger \cosh r- a \sinh r)(a \cosh r- a^\dagger \sinh r),
\end{equation} 
we obtain the total average photon number in all four modes as
\begin{equation}
\nonumber
\bar{n}
=\langle \hat{n}_1+\hat{n}_2+\hat{n}_3+\hat{n}_4\rangle=2\cosh 2r+4\sinh^2 r.
\end{equation}
In this scheme, only specific values of $r$ result in the improvement of the success probability for the Bell discrimination. The best suggested value in Ref.~\cite{sZaidi2013} is  $r=0.6585$ with which the average photon number in the Bell state after the squeezing is $\bar{n}=6.00029$; this results in the success probability of the Bell measurement $0.643$. 

(iv) The scheme proposed by Ewert and van Loock \cite{sEwert2014} employs ancillary multi-photon entanglement that is similar to the one used in Grice's scheme~\cite{sGrice2011} 
to increase the success probability. The total number of photons that go into the Bell measurement setup can be counted as $\bar{n}=4N_m+2$ where $N_m$ is the number of ancillary states, which yields the success probability $P_s=1-2^{-{N_m}-1}$. Thus, we can rewrite the success probability with respect to the average photon usages $\bar{n}$ as $P_s=1-2^{-{\bar{n}/4}-1/2}$.

We plot $P_s$ against $\bar{n}$ for all above-mentioned schemes in Fig.~2 of the main Letter which shows that our scheme shows the best performance.

\section{Error probabilities in lossy environment}

The evolution of optical qubits in a lossy environment can be described by solving the master equation,
\begin{equation}
\nonumber
\frac{d\rho}{dt}=\gamma(\hat{J}+\hat{L})\rho,
\end{equation}
where $\hat{J}\rho=\sum_i\hat{a}_i\rho\hat{a}_i^{\dag}$,
$\hat{L}\rho=-\sum_i(\hat{a}^{\dag}_i\hat{a}_i\rho+\rho\hat{a}^{\dag}_i\hat{a}_i)/2$,
and $\hat{a} (\hat{a}^{\dag})$ is the annihilation (creation)
operator for the $i$-th mode, and $\gamma$ is the decay constant.
 Here the loss rate is given by $\eta=
1-e^{-\gamma t}$. Thus a logical qubit with arbitrary $N$ photons 
$\ket{\psi^{(N)}}=a\ket{+}^{\otimes N}+b\ket{-}^{\otimes N}$ evolves to
a mixed state as
\begin{equation}
\begin{aligned}
\label{eq:master}
\nonumber
&\ket{\psi^{(N)}}\xrightarrow{\eta}(1-\eta)^N\ket{\psi^{(N)}}\bra{\psi^{(N)}}\\
&+\frac{1}{2}\sum^{N}_{k=1}\binom{N}{k}(1-\eta)^{N-k}\eta^{k}\\
&~~~~~~\times\bigg(\ket{\psi^{(N-k)}}\bra{\psi^{(N-k)}}+\ket{\psi_{-}^{(N-k)}}\bra{\psi_{-}^{(N-k)}}\bigg),
\end{aligned}
\end{equation}
where $\binom{N}{k}$ represents the binomial coefficient. As we can see here, losses in a logical qubit occur with rate $P=1-(1-\eta)^{N}$. When $k$ photons are lost, the possible resulting state of the qubit is either $\ket{\psi^{(N-k)}}=a\ket{+}^{\otimes N-k}+b\ket{-}^{\otimes N-k}$ or $\ket{\psi_{-}^{(N-k)}}=a\ket{+}^{\otimes N-k}-b\ket{-}^{\otimes N-k}$. Here the former contains no logical error, while the latter contains a Pauli-$Z$ error. Therefore, if a photon is lost, a qubit experiences a Pauli-$Z$ error with probability $1/2$. 

Likewise, we can calculate the failure probability, $1-P_s$, of the Bell measurement in a lossy environment as
\begin{equation}
\begin{aligned}
\label{eq:master}
\nonumber
P_f(\eta)=\sum^{N}_{k=0}\binom{N}{k}(1-\eta)^{N-k}\eta^{k}\left(\frac{1}{2}\right)^{N-k}=\left(\frac{1+\eta}{2}\right)^N,
\end{aligned}
\end{equation} 
where $(1/2)^{N-k}$ is the failure rate obtained when k photons are lost.

\section{Telecorrector circuit and noise thresholds for fault-tolerant quantum computing}

The telecorrector circuit is composed of CZ, Hadamard, $\ket{+}$ state, and $X$-basis measurement \cite{sLund2008}.  For the lowest level of concatenation, the errors can be modeled as follows: When Hadamard or CZ gates fail, the teleported qubit is assumed to experience depolarization, modeled by a random Pauli operation applied to the qubit, {\it i.e.} $Z$ and $X$ errors occur independently with the equal probability $1/2$. If a loss occurs in quantum memory or gate operations, the qubit experiences a Pauli $Z$ error with probability $1/2$. For higher levels of concatenation, we use the same error models described in Ref.~\cite{sDawson2006}. Based on these, we performed a series of Monte Carlo simulation (using C++) to obtain the corrected error rates for a range of $\eta$ with different $N$. The resulting error rates in a lower level are used for the next level of error correction. If the error rates tend to zero in the limit of many levels of concatenation, fault-tolerant quantum computation is possible with those certain $\eta$ and $N$. In this way, the noise thresholds of our model are obtained as presented in the main Letter.


\begin{thebibliography}{99}
\bibitem{PKok2007} P. Kok, W. J. Munro, K. Nemoto, T. C. Ralph, J. P. Dowling, and G. J. Milburn,  Rev. Mod. Phys. {\bf 79}, 135 (2007).
\bibitem{Ralph2010} T. C. Ralph and G. J. Pryde, Progress in Optics {\bf 54}, 209 (2010).
\bibitem{Knill2001} E. Knill, R. Laflamme, and G. J. Milburn, Nature {\bf 409}, 46 (2001).

\bibitem{Lut99} N. L\"utkenhaus, J. Calsamiglia, and K.-A. Suominen, Phys. Rev. A {\bf 59}, 3295 (1999).

\bibitem{Calsa2001} J. Calsamiglia and N. L\"{u}tkenhaus, App. Phys. B {\bf 72}, 67 (2001).

\bibitem{Grice2011} W. P. Grice, Phys. Rev A {\bf 84}, 042331 (2011).
\bibitem{Ewert2014} F. Ewert and P. van Loock, Phys. Rev. Lett. {\bf 113}, 140403 (2014).
\bibitem{Zaidi2013} H. A. Zaidi and P. van Loock, Phys. Rev. Lett. {\bf 110}, 260501 (2013).

\bibitem{Jeong2001} H. Jeong, M. S. Kim, and J. Lee, Phys. Rev. A {\bf 64}, 052308 (2001);
H. Jeong and M. S. Kim, Quantum Information and Computation {\bf 2}, 208 (2002).

\bibitem{SWLEE13} S.-W. Lee and H. Jeong, Phys. Rev. A  {\bf 87}, 022326 (2013).


\bibitem{Bennett93} C. H. Bennett, G. Brassard, C. Crepeau, R. Jozsa, A. Peres, and W. K. Wootters, Phys. Rev. Lett. {\bf 70}, 1895 (1993).



\bibitem{Gottesman1999} D. Gottesman and I. L. Chuang, Nature {\bf 402}, 390 (1999).

\bibitem{JWPan2012} J.-W. Pan, Z.-B. Chen, C.-Y. Lu, H. Weinfurter, A. Zeilinger, and M. Zukowski, Rev. Mod. Phys. {\bf 84}, 777 (2012).
\bibitem{Yao2012} X.-C. Yao, T.-X. Wang, P. Xu, H. Lu, G.-S. Pan, X.-H. Bao, C.-Z. Peng, C.-Y. Lu, Y.-A. Chen and J.-W. Pan, Nature Photonics {\bf 6}, 225 (2012).
\bibitem{Huang2011} Y.-F. Huang, B.-H. Liu, L. Peng, Y.-H. Li, L. Li, C.-F. Li, and G.-C. Guo, Nature Communications {\bf 2}, 546 (2011).
\bibitem{Yao2012Nature} X.-C. Yao, T.-X. Wang, H.-Z. Chen, W.-B. Gao, A. G. Fowler, R. Raussendorf, Z.-B. Chen, N.-L. Liu, C.-Y. Lu, Y.-J. Deng, Y.-A. Chen, and J.-W. Pan, Nature {\bf 482}, 489 (2012).

\bibitem{deterministic}
C. Sch\"on, E. Solano, F. Verstraete, J. I. Cirac, and M. M. Wolf, Phys. Rev. Lett.  {\bf 95}, 110503 (2005).

\bibitem{Lindener2009} N. H. Lindner and T. Rudolph, Phys. Rev. Lett. {\bf103}, 113602 (2009).
\bibitem{Young2006} R. J. Young, R. M. Stevenson, P. Atkinson, K. Cooper, D. A. Ritchie, A. J. Shields, New J. Phys. {\bf 8},  29 (2006).


\bibitem{Phoenix90} S. J. D. Phoenix,  Phys. Rev. A {\bf41}, 5132 (1990).

\bibitem{Shor1996} P. W. Shor, in Proceedings of the 37th Annual Symposium on Fundamentals of Computer Science (IEEE Computer Society Press, Los Alamotios, CA, 1996), pp. 56-65.

\bibitem{Steane1996} A. M. Steane, Phys. Rev. A {\bf 54}, 4741 (1996).

\bibitem{Dawson2006} C. M. Dawson, H. L. Haselgrove, and M. A. Nielsen, Phys. Rev. A {\bf 73}, 052306 (2006).

\bibitem{Lund2008} A. P. Lund, T. C. Ralph, and H. L. Haselgrove, Phys. Rev. Lett. {\bf100}, 030503 (2008).
\bibitem{coherent_1}
 H. Jeong and M. S. Kim, Phys. Rev. A {\bf 65}, 042305 (2002).

\bibitem{coherent_2} 
T.C. Ralph, A. Gilchrist, G.J. Milburn, W.J. Munro, S. Glancy, Phys. Rev. A {\bf 68}, 042319 (2003).

\bibitem{Ralph2005} T. C. Ralph, A. J. F. Hayes, and A. Gilchrist, Phys. Rev. Lett. {\bf 95}, 100501 (2005).
\bibitem{Hayes2010} A. J. F. Hayes, H. L. Haselgrove, A. Gilchrist, and T. C. Ralph, Phys. Rev. A {\bf 82}, 022323 (2010).

\bibitem{Sean2010} S. D. Barrett and T. M. Stace, Phys. Rev. Lett. {\bf 105}, 200502 (2010).

\bibitem{ex1} A. Zavatta, M. D'Angelo, V. Parigi, and M. Bellini,
Phys. Rev. Lett. {\bf 96}, 020502 (2006).

\bibitem{ex2} M. D'Angelo, A. Zavatta, V. Parigi, and
M. Bellini, J. Mod. Opt. {\bf 53}, 2259 (2006).

\bibitem{ex3}
H. Jeong, A. Zavatta, M. Kang, S.-W. Lee, L. S. Costanzo, S. Grandi, T. C. Ralph, and M. Bellini, Nature Photonics {\bf 8},  564 (2014).

\end{thebibliography}

\begin{thebibliography}{99}
\bibitem{sGrice2011} W. P. Grice, \pra {\bf 84}, 042331 (2011).
\bibitem{sZaidi2013} H. A. Zaidi and P. van Loock, \prl {\bf 110}, 260501 (2013).
\bibitem{sEwert2014} F. Ewert and P. van Loock, Phys. Rev. Lett.113, 140403 (2014).
\bibitem{sLund2008} A. P. Lund, T. C. Ralph, and H. L. Haselgrove, \prl {\bf100}, 030503 (2008).
\bibitem{sDawson2006} C. M. Dawson, H. L. Haselgrove, and M. A. Nielsen, \pra {\bf73}, 052306 (2006).
\end{thebibliography}
\end{document}